\documentclass[aps,prb,twocolumn,citeautoscript]{revtex4-1}        
\usepackage{amsmath,amssymb,mathrsfs,bm,feynmf,setspace,xspace,float}
\usepackage{graphicx}    
\usepackage[tight]{subfigure}  
\usepackage{color} 
\usepackage[colorlinks=true]{hyperref} 
\hypersetup{
    bookmarks=true,         
    unicode=false,          
    pdftoolbar=true,        
    pdfmenubar=true,        
    pdffitwindow=false,     
    pdfstartview={FitH},    
    pdftitle={My title},    
    pdfauthor={Author},     
    pdfsubject={Subject},   
    pdfcreator={Creator},   
    pdfproducer={Producer}, 
    pdfkeywords={keyword1} {key2} {key3}, 
    pdfnewwindow=true,      
    colorlinks=true,       
    linkcolor=magenta, 
    citecolor=blue,        
    filecolor=magenta,      
    urlcolor=cyan           
} 
\newcommand{\al}{\alpha}  
 
\newcommand{\g}{\gamma}
 
\newcommand{\e}{\epsilon}

\newcommand{\la}{\lambda}
\newcommand{\La}{\Lambda}

\newcommand{\s}{\sigma}

\newcommand{\w}{\omega}

\newcommand{\G}{\Gamma}
\renewcommand{\S}{\Sigma}

\newcommand{\pd}{\partial}

\newcommand{\bra}[1]{\left\langle #1\right|}
\newcommand{\ket}[1]{\left| #1\right\rangle}

\newcommand{\beq}{\begin{equation}}
\newcommand{\eeq}{\end{equation}}
\newcommand{\Beq}{\begin{eqnarray}}
\newcommand{\Eeq}{\end{eqnarray}}
\newcommand{\bml}{\begin{multline}}

\newcommand{\eeqm}{\end{multline}}

\newcommand{\bsp}{\begin{split}}
\newcommand{\esp}{\end{split}}
\newcommand{\down}{\downarrow}
\newcommand{\up}{\uparrow}

\renewcommand{\b}[1]{{\bm #1}}
\renewcommand{\t}{\tilde}

\newcommand{\inv}{^{-1}}

\newcommand{\mc}{\mathcal}

\renewcommand{\t}{\tilde}
\newcommand{\ra}{\rightarrow}
\newcommand{\req}[1]{Eq.\thinspace{}(\ref{eq:#1})}

\newcommand{\rfig}[1]{Fig.\thinspace{}\ref{fig:#1}}

\newcommand{\ts}{\thinspace{}}

\DeclareMathOperator{\Tr}{Tr}

\DeclareMathOperator{\sgn}{sgn}
 
\DeclareMathOperator{\im}{Im}

\newcommand{\ie}{i.e.\@\xspace}
\newcommand{\ns}{\@\xspace}
\newcommand{\bc}{\b{{\mc A}}}

\begin{document} 

\title{Interacting Weyl semimetals: characterization via the topological Hamiltonian\\ and its breakdown}  
 \author{William Witczak-Krempa$^1$}
 \author{Michael Knap$^{2,3}$} 
 \author{Dmitry Abanin$^1$}
 \affiliation{$^1$Perimeter Institute for Theoretical Physics, Waterloo, Ontario N2L 2Y5, Canada \\
$^2$Department of Physics, Harvard University, Cambridge, MA, 02138, USA \\ 
$^3$ITAMP, Harvard-Smithsonian Center for Astrophysics, Cambridge, MA 02138, USA}

  \date{\today}
\begin{abstract} 
Weyl semimetals (WSMs) constitute a 3D phase with linearly-dispersing Weyl excitations at low energy, 
which lead to unusual electrodynamic responses and
open Fermi arcs on boundaries. We derive a simple criterion to identify and characterize 
WSMs in an interacting setting using the exact electronic Green's function at zero frequency, which defines a 
topological Bloch Hamiltonian. We apply this criterion   
by numerically analyzing, via cluster and other methods, interacting lattice models with and without time-reversal symmetry. 
We identify various mechanisms for how interactions move and renormalize Weyl fermions.   
Our methods remain valid in the presence of long-ranged Coulomb repulsion.
Finally, we introduce a WSM-like phase for
which our criterion breaks down due to fractionalization: the charge-carrying 
Weyl quasiparticles are orthogonal to the electron.   
\end{abstract}  
\maketitle  
 
The emergence of (quasi)relativistic excitations in quantum condensed matter has stimulated
much theoretical and experimental research, especially following the discoveries of graphene\cite{geim,castro} 
and 3D topological 
insulators\cite{RMP_TI,RMP_TI2}, both of which host
2D massless Dirac fermions. 
More recently, a 3D analog of graphene, the Weyl semimetal (WSM), has piqued physicists' curiosity,  
partially due to its potential for realization in transition metal oxides with strong interactions
and spin-orbit coupling\cite{wan,fang,will-pyro1,will-rev}, or heterostructures\cite{burkov}.   
Such a phase has stable massless Weyl quasiparticles, which can be viewed as half-Dirac fermions. 
These lead to unique open Fermi arc surface states\cite{wan} and electromagnetic 
responses\cite{nielsen,ahe_haldane,ran,turner,vafek,sid,hosur_friedel,drew,panfilov}. 
Such properties rely
on the topological nature of the Weyl points\cite{volovik-book}, which are monopoles of the non-interacting Berry curvature.    
As WSMs naturally arise in interacting lattice models\cite{turner,will-rev,vafek}, it is important to characterize them
without relying on free-electron or  
field-theoretic approaches\cite{ara-pyro1,wang_gapless,mastropietro,sekine},   
neither of which is sufficient to provide accurate predictions for most realistic systems.  
Moreover, an efficient method for searching for Weyl points in the interacting setting is desired 
because they generally occur at incommensurate points in the Brillouin Zone (BZ), often due to
spontaneous symmetry breaking. 
 
We provide a simple criterion to identify and characterize WSMs in the quantum many-body setting  
based on the electronic lattice Green's function.  
Specifically, we use an effective Bloch Hamiltonian (dubbed ``topological Hamiltonian''\footnote{This notation
was introduced in Ref.\ts\onlinecite{wang_Ht} in the context of interacting topological insulators.
We keep it, although we deal with gapless systems. This is motivated because
the topological Hamiltonian captures the Berry phase properties of the electrons (such as hedgehogs).}) defined from the zero-frequency many-body    
Green's function, and argue that its eigenstates retain the Berry phase properties of the Weyl nodes.  
This allows for the extraction of the non-trivial surface states\cite{wan} 
and anomalous quantum Hall (AQH) response\cite{ahe_haldane,ran} of interacting WSMs.
We apply our results in conjunction with Cluster Perturbation Theory\cite{senechal} to study the physics
of two interacting lattice models for WSMs, unraveling diverse interaction effects 
on the renormalization of the Weyl points. We also discuss the effects of long-range Coulomb repulsion  
which marginally destroys the quasiparticles\cite{abrikosov}, and argue that our approach 
remains valid in that case.  
Finally, we provide an instance where such methods break down due to a simple fractionalization 
into an orthogonal\cite{orthog} WSM. 
Our analysis naturally relates to previous works that characterized interacting topological insulators
\cite{Wang2012,wang_simple,ara-pyro1,hung13,lang,deng,laubach,grandi} by means of the many-body Green's function and associated Berry curvature,
 but differs in the sense that we study gapless systems.    

{\bf Characterizing interacting Weyl semimetals}: Non-interacting WSMs have
a Fermi surface consisting of a finite number of points in the BZ, at which 2 bands meet linearly.  
Each such Weyl point can be identified with a hedgehog singularity of the Berry 
curvature, $\nabla\times \b a(\b k)$, 
\ie a monopole of this $k$-space ``magnetic'' field. Here, $\b a$ is the Berry connection defined via the
occupied Bloch states. 
Knowledge of this monopole structure naturally leads to a description of the unusual open Fermi arc surface states\cite{wan}, 
and AQH response\cite{ahe_haldane,ran}. 
In the presence of interactions that inevitably arise in realistic systems,  
the above band structure description no longer applies. However, we demonstrate that the essential features 
of the WSM remain robust, and can be understood in terms of the
zero-frequency Green's function.  

We focus on short range interactions, while the effects of the 
long-ranged Coulomb repulsion are discussed towards the end. 
The central tool in our analysis is the imaginary-frequency Green's function, 
$G(i\w,\b k)$. It is a matrix in 
spin/orbital/sublattice space, and $\b k$ belongs to the BZ of the lattice of the interacting system. 
A key observation is that one can define a many-body Berry connection $\b{{\mc A}}(\b k)$, and associated  
Berry curvature $\nabla\times \b{{\mc A}}$, using the zero-frequency Green's function. One begins by defining
the so-called topological Hamiltonian:
\begin{align}
  \mc H_t(\b k)=-G(0,\b k)\inv=\mc H(\b k)+\S(0,\b k)\,, 
\end{align}
where $\mc H$ is the Bloch Hamiltonian of the non-interacting system, while $\S(i\w,\b k)$ is the exact self-energy matrix.
$\mc H_t$ plays the role of an effective Bloch Hamiltonian: its eigenstates can be loosely viewed as  
substitutes of the Bloch states of the 
non-interacting system. The many-body Berry connection can then be introduced in exact analogy with non-interacting systems:  
  $\b{{\mc A}}(\b k)=-i\sum_{\rm R-zeros}\langle n\b k |\nabla | n\b k\rangle  \,,$ \label{eq:A}
where $\mc H_t(\b k)|n\b k\rangle=\t\xi_n(\b k)|n\b k\rangle$ and $\{\t\xi_n(\b k)\}$ defines
the band structure of $\mc H_t$. R-zero\cite{Wang2012} signifies an eigenstate  
with $\t\xi_n(\b k)\leq 0$. In the non-interacting limit, R-zeros reduce to occupied states, and $\b{{\mc A}}$ to $\b a$.  
We now argue that Weyl points of the interacting system can then be identified with monopoles of $\nabla\times\bc$
(analogously for higher charge monopoles\cite{multi-weyl}).    
An equivalent but more practical criterion follows: an interacting system is a WSM if the
band structure of the topological Hamiltonian $\mc H_t$ has Weyl nodes 
at the Fermi level, which identify 
the Weyl nodes of the interacting system.  

To understand the above criterion, let us consider a non-interacting WSM for which short-ranged interactions 
(attractive or repulsive) are adiabatically turned on.
The latter are irrelevant in the renormalization group sense, \ie at low energy, and one thus obtains a 
\emph{Weyl liquid}, where excitations have an infinite lifetime only on the Fermi surface, \ie at the Weyl nodes. 
By adiabacity, the monopole structure of the non-interacting Green's function cannot be destroyed in the Weyl liquid.
The many-body Berry connection $\bc$ captures the monopole of Berry flux\cite{wang_gapless} 
associated with the Weyl quasiparticles.   
This relates to Haldane's statement\cite{ahe_haldane} about using the Berry curvature of the quasiparticles 
of a Fermi liquid to
determine its AQH response (which translates to our expression for the latter, \req{K}, being valid in that case), as one can approach a Weyl liquid from its parent Fermi liquid by tuning the doping. 

We now support the above arguments by deriving the AQH response of a WSM in terms of the  
generalized Berry curvature $\bc$.  
We proceed by evaluating the many-body Chern number for 2D surfaces away from the Weyl points in the BZ.\cite{wang_gapless}  
More precisely, we will show that the anomalous part of the Hall conductivity reads:
\begin{align}
  \s_{ab}=\frac{e^2}{2\pi h}\e_{abc}K^c\,; \qquad 
  \b K =\int_{\rm BZ} \frac{d^3\b k}{2\pi}\, \nabla \times \bc (\b k) \,, \label{eq:K} 
\end{align}
where $\e_{abc}$ is the Levi-Civita tensor. 
\req{K} generalizes the non-interacting formula\cite{ahe_haldane}, and 
can be collapsed to Fermi surface data: $\b K=\sum_m q_m \b k_m$, where $\b k_m$ is a Weyl node 
of the interacting system, and $q_m=\pm 1$, its monopole charge.  
\req{K} can be deduced by starting with the frequency-dependent Green's function. 
For simplicity, we consider a fixed $k_x$ away from the Fermi surface of the interacting WSM. It follows that
$G(i\w,\b k)$ defines a gapped 2D Green's function in the $k_{y,z}$ plane. We can compute the many-body Chern number
associated with $G$ at fixed $k_x$\cite{so,ishikawa_86}:   
\begin{align}
 \!\!\!\! C_{x}(k_x)\!=\!\!\int\!\! \frac{d\w d\b k_{y,z}}{24\pi^2}\e_{\mu\nu\rho x}\!
   \Tr G\pd_\mu G\inv G\pd_\nu G\inv G\pd_\rho G\inv \label{eq:chern}  
\end{align}
The $x$-component of the anomalous Hall vector is then the integral over the Chern number: 
$K_x=\int dk_x C_x(k_x)$.   
We note that this latter expression agrees with the so-called Adler-Bell-Jackiw anomaly coefficient of the 
current correlator (see Appendix~\ref{ap:ahe}). 
Now, to recover \req{K}, we adiabatically deform the interacting Green's function 
into the topological Green's function, $G_t(i\w,\b k)\inv=i\w-\mc H_t(\b k)$, via the interpolation:
$g_\la(i\w,\b k)=(1-\la)G(i\w,\b k) + \la G_t(i\w,\b k)$, $0\leq\la\leq 1$. Indeed, for any slice 
away from the Fermi surface, the gap of $g_\la$ remains open during the protocol 
since $g_\la(0,\b k)=G(0,\b k)$ for all $\la$. Further, $g_\la(i\w,\b k)$ does not have zero eigenvalues\cite{wang_simple}.
Thus, the many-body Chern number cannot change  
as $\la$ varies from 0 to 1, being a topological index, and we can use $g_{\la=1}=G_t(i\w,\b k)$ to compute $C_x$. The frequency
integral then yields \req{K} (Appendix~\ref{ap:ahe}).    

{\bf Using topological Hamiltonians numerically}: We study two lattice models of interacting WSMs
numerically to show the usefulness of the topological Hamiltonian approach. We identify and explain the motion
and renormalization of the Weyl points as a function of the interaction strength.
We consider Hubbard models:
\begin{align}
  H = H_0+U\sum_r n_{r,\up}n_{r,\down} -\mu \sum_{r,\s}n_{r,\s}\, ,
\end{align}
where $H_0$ is a tight-binding hopping Hamiltonian of spin-1/2 electrons, which are created at site $r$ 
by $c_{r,\s}^\dag$, and their number density per spin projection is $n_{r,\s}=c_{r,\s}^\dag c_{r,\s}$. 
$U$ is the Hubbard interaction parameter; we consider both the attractive and
repulsive cases. We study models that are defined on the cubic lattice, have particle-hole symmetry and  
are WSMs at the non-interacting level, 
and fix the chemical potential at the nodes, $\mu=U/2$. 

\begin{figure}
        \centering
        \includegraphics[width=0.45\textwidth]{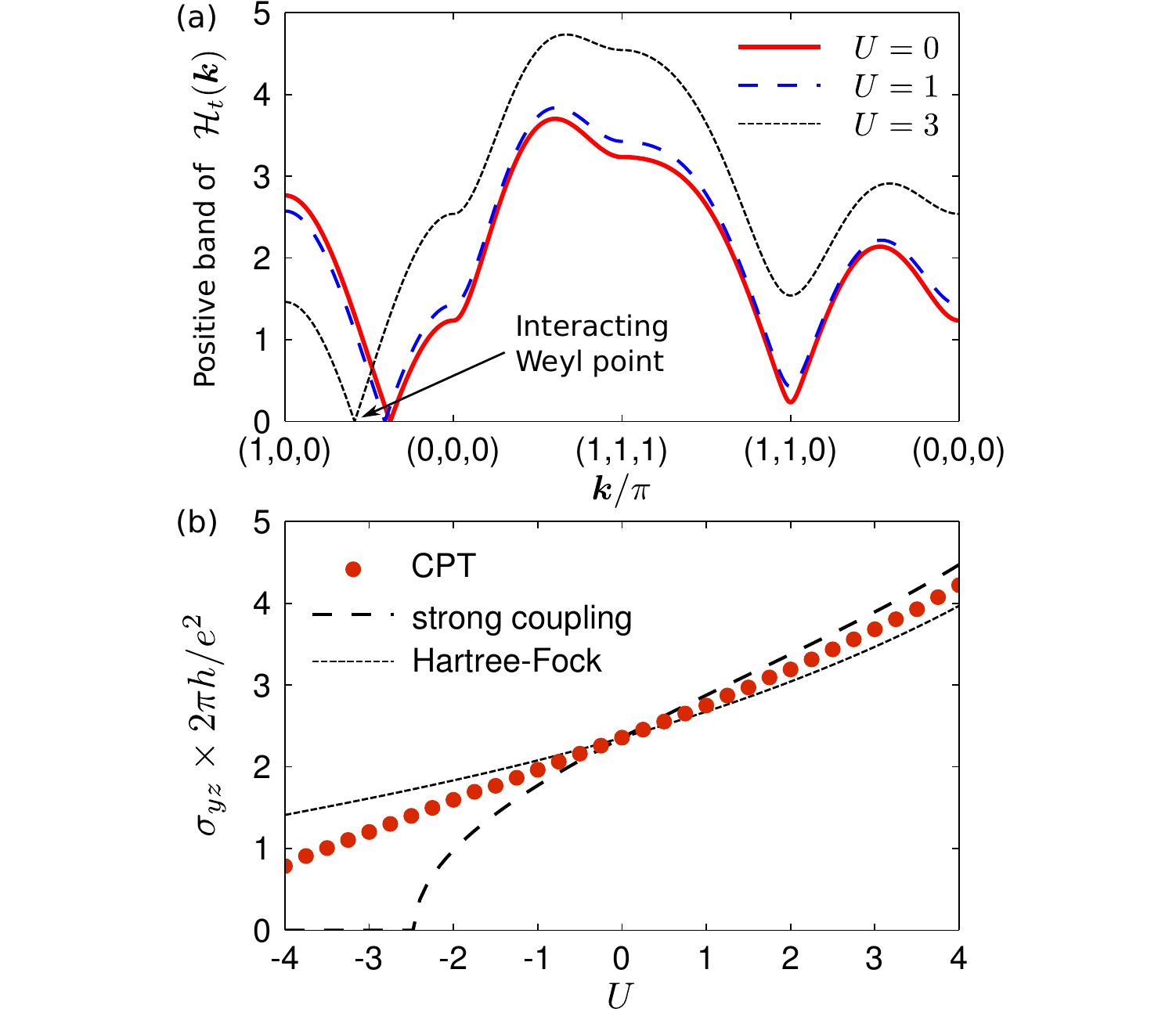}
        \caption{ \label{fig:hEff}
        a) Positive band of the topological Hamiltonian $\mc H_t=-G(0,\b k)^{-1}$ of model I with
        varying interaction strength $U$ along a cut through the BZ. b)   
        The Weyl points move with varying $U$, altering the
        Hall conductivity $\s_{yz}$. Filled circles come from the numerical simulations.  
        $\s_{yz}$ in the strong/weak coupling limit is shown (dashed/dotted line).  
        The numerical results are obtained with Cluster Perturbation Theory for a cluster of size 
        $N_c=2^3$; the single-particle Hamiltonian has $m=3/2,\,     
        k_0=3\pi/8$. }        
\end{figure}

Model I breaks time-reversal and is defined by\cite{ran}:
\begin{multline}
 \!\!\! H_0\!=\!\sum_{\b k} c_{\b k}^\dag \big[\{2t(\cos k_x-\cos k_0) 
+m(2-\cos k_y-\cos k_z)\}\s_x \\ +2t\sin k_y\,\s_y+2t\sin k_z\,\s_z\big]c_{\b k} \,, \label{eq:modelI}
\end{multline}
where the spacing of the cubic lattice has been set to unity, and 
the fermion operators are vectors in spin space. The Pauli matrices $\s_a$ 
act on the latter. Below we set $t=1$. 
Depending on the parameters $k_0$ and $m$, $H_0$ can have 2, 6 or 8 Weyl nodes. We focus on the regime where
it only has 2 nodes, located on the BZ boundary at $\b k=\pm(k_0,0,0)$. See \rfig{hEff}a for the $U=0$
band structure (recall that in that limit $\mc H_t=\mc H$). The anomalous Hall vector is thus given by 
$\b K=2k_0\hat x$, \ie $\s_{yz}=(e^2/2\pi h)2k_0$.  

We now turn to the study of the interacting Hamiltonian using Cluster Perturbation Theory\cite{senechal} (CPT). 
This method, which is related to Dynamical Mean Field Theory, allows for an efficient
numerical analysis. In essence, one first decomposes the periodic system into clusters with $N_c$ sites. Exact
diagonalization is used to obtain the exact cluster Green's function. The Green's function of the lattice system
is then obtained via strong-coupling perturbation theory. 
CPT becomes exact in the limit $U\ra 0$ and at strong coupling, $U\ra\infty$; 
it is controlled in the sense that convergence can be monitored with increasing the cluster size.
We emphasize that it is not perturbative in $U$. See Appendix~\ref{ap:cpt} for more details. 

CPT allows a direct evaluation of the topological Hamiltonian $\mc H_t$, so that we can easily track
the location of the Weyl points of the interacting system as a function of $U$. 
The results we present are for clusters of size $N_c=2^3$, at which point  
reasonable convergence with $N_c$ has been achieved (see Appendix~\ref{ap:cpt} for further information
regarding the convergence). A further increase of $N_c$ would not affect our conclusions.
We set $m=3/2$, and $k_0=3\pi/8$. The 
positive band of the topological Hamiltonian is shown in \rfig{hEff}a for a cut 
through the BZ and for different $U$ values.  
With increasing $U>0$, the Weyl points move to larger magnitude of the wave vector.
This directly corresponds to an increase of the Hall conductivity $\sigma_{yz}$, 
\req{K}, as shown in \rfig{hEff}b. The red circles are evaluated  
numerically with CPT.  
We also show the analytic strong coupling result (perturbative in $1/U$; derived in
Appendix~\ref{ap:models}), \ie for single-site clusters, which captures the 
overall trend. 
For attractive interactions $U<0$, 
the trend is opposite: the Weyl points move towards $\b k=0$. 
One can understand this heuristically: a positive/negative $U$ enhances/reduces the ferromagnetic moment 
$\langle c_r^\dag\s_x c_r\rangle$ (already present at $U=0$), thus enhancing/reducing the  
Hall conductivity. A crude estimate of this effect can be obtained using mean field theory
(Appendix~\ref{ap:models}),
as shown in \rfig{hEff}b.  

\begin{figure*}[t!]
        \centering
        \includegraphics[width=0.95\textwidth]{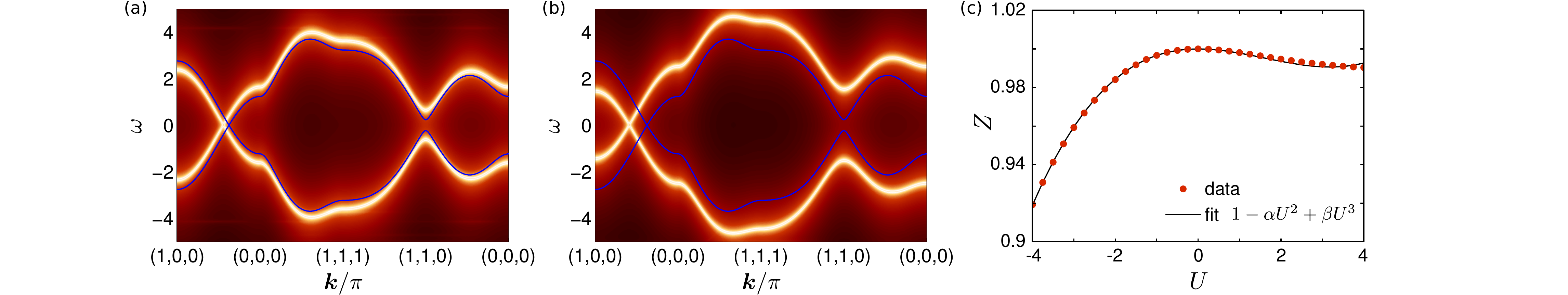}   
        \caption{ Density plot of the single particle spectral function $A(\omega,\b k)$ for a) $U=1$,  
          b) $U=3$ obtained via CPT, shown on a logarithmic color scale. 
          The non-interacting band structure (solid blue line) is the 
          same as in \rfig{hEff}. c) Dependence of the residue $Z$ of the Weyl quasiparticles on $U$; it is 
          well approximated by a cubic polynomial.} 
        \label{fig:A}
\end{figure*}   
 
In studying the Weyl points and AQH response of the many-body system, the topological Hamiltonian 
allowed a streamlined analysis by circumventing the need for the full frequency-dependent
Green's function. We now discuss some of the properties arising from the latter but not captured by $\mc H_t$. 
The spectral function $A(\omega,\b k)=-\Tr\im G_R(\omega+i0^+,\b k)/\pi$ obtained using CPT 
for $U\geq 0$ is shown in \rfig{A}. The   
linearly dispersing Weyl modes
can clearly be seen. In the interacting WSM only the excitations at the Weyl points 
remain sharp. The scattering rate of an excitation with momentum exactly at a Weyl point and with small
frequency vanishes like $|\w|^5$, as can be obtained perturbatively as shown in 
Appendix~\ref{ap:liquid}. 
This is smaller than the Fermi liquid result $\w^2$, owing to the vanishing 
density of states at the Fermi level in a WSM. 
As in a FL, the weight of the quasiparticles $Z$ will be reduced with increasing interactions. 
(When the Weyl nodes are related by symmetry they share the same $Z$, which is the case in this work.) 
The result is plotted in \rfig{A}c, and as expected behaves as $Z\approx 1-\al U^2$ at small $U$, $\al>0$. 
\begin{figure}
        \centering
        \includegraphics[width=0.46\textwidth]{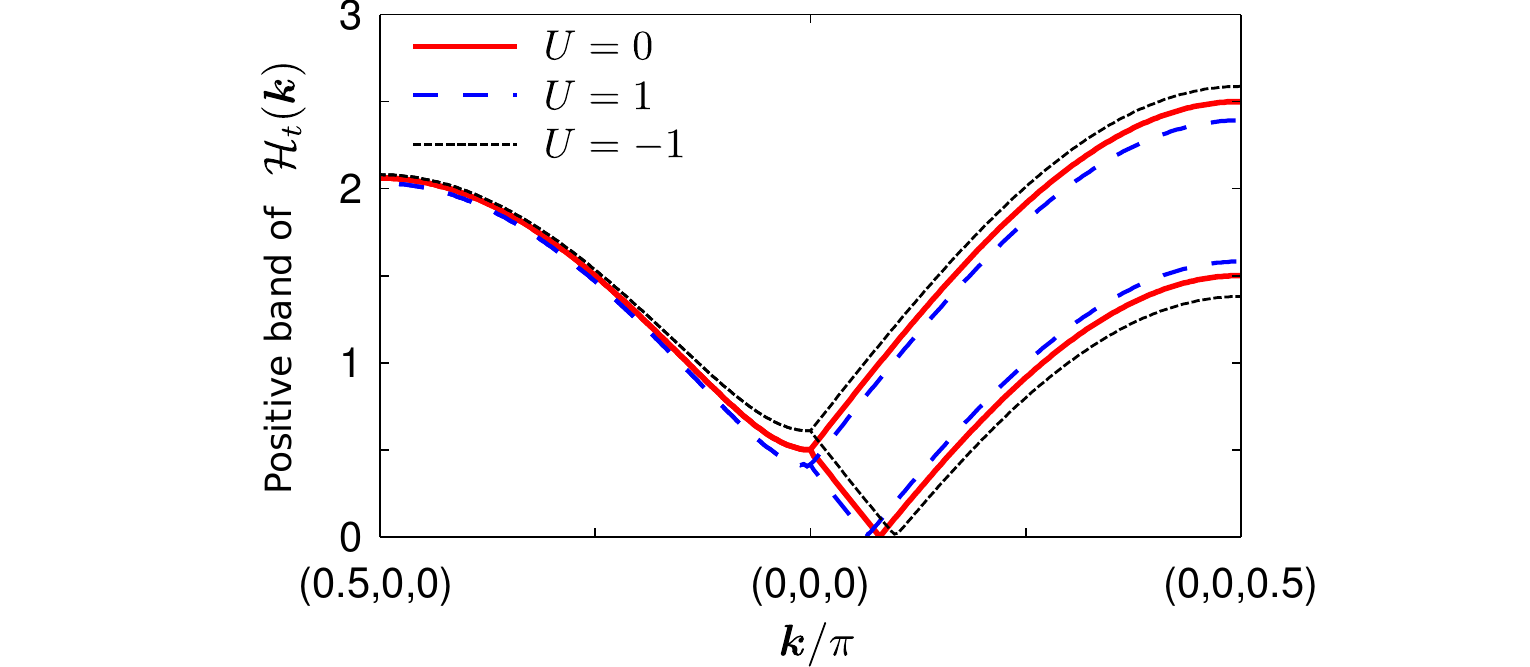}  
        \caption{ Positive band of the topological Hamiltonian $\mc H_t$ of model II ($\e=0.5$) with varying
        Hubbard $U$ along a cut through the BZ. $U=0$ is the non-interacting band structure.} 
        \label{fig:heff2} 
\end{figure}   

We introduce a new model which, in contrast to model I, preserves TRS but not inversion, 
and as such is a representative of the second family of WSMs. We show that the influence of interactions 
on the motion of the Weyl points has an altogether different physical origin as compared to model I, 
but a connection can be made by interchanging the role of magnetic and charge orders.
The tight-binding Hamiltonian of model II reads:  
\begin{align}
  \!H_0\!=2t\!\sum_{\b k,b=x,y,z}\! c_{\b k}^\dag \, \s_b\sin k_b \, c_{\b k}
  + \e H_{\rm cdw}\,, \label{eq:modelII}  
\end{align}
where $H_{\rm cdw}$ corresponds to a $(\pi,\pi,0)$ charge density wave (CDW) on the cubic lattice
where the chemical potential is staggered by  $\pm \e$ in a checkerboard fashion in the $xy$ plane.  
When $\e=0$, we do not expect the 8 Weyl points to move 
under the effect of interactions (modulo possible instabilities\cite{joseph} beyond a critical $U$) 
because they are located at special high-symmetry $k$-points. The CPT calculation  
corroborates this. We thus need to turn on a finite $\e$ to get non-trivial evolution. At $U=0$, we find 
a total of 16 Weyl points when $|\e|<1$, setting $t=1$. 
(Going from 8 to 16 Weyl points as $\e$ is turned on does not violate  
the indivisible nature of Weyl points since the CDW changes the BZ.)
When $\e=0$, four Weyl points occur at $k_z=0$, while four other ones at $\pi/a$, where we have 
reinstated the spacing of the original cubic lattice $a$. When $0<\e< 1$, the eight nodes at $k_z=0$ ``split'' 
to ones at $k_za=\pm \sin\inv(\e/2)$, similarly for $k_z=\pi/a$. A finite $U$ moves the eight nodes nearest to $k_z=0$
towards/away from $k_z=0$ since a repulsive/attractive $U$ disfavors/favors the charge imbalance. This 
is confirmed by \rfig{heff2},
which shows $\mc H_t$ obtained using CPT.

{\bf Long-ranged Coulomb interaction}: 
We have so far limited our discussion to short-ranged interactions. However, in an electronic
WSM the screening of the Coulomb interaction is weak due to the vanishing density of
states at the Fermi energy. Using RPA, it was shown\cite{abrikosov} that for linearly dispersing
electrons in 3D interacting via an instantaneous Coulomb $1/r$ repulsion, the quasiparticle
at the node $\b k_0$ is marginally destroyed: $\im\S_R(\w+i0^+,\b k_0)\sim |\w|$, resulting in a
``marginal Weyl liquid''. Notwithstanding, this does not alter the fundamental Berry curvature
structure around the (marginal) Weyl point. Indeed, let us consider the low-energy description near such
an isotropic point: $\mc H_t(\b k)=-G(0,\b k)\inv=f(k)\b k\cdot\b \s$, where $f=1+\la\ln(\La/k)$\cite{abrikosov}. 
Crucially, the Berry curvature $\nabla\times \b{{\mc A}}$ is independent of the overall real   
renormalization factor $f$ as it measures the \emph{complex phase} of the $G$ eigenstates as they 
are parallel transported in the BZ. Thus, the Berry flux through a small sphere 
surrounding the Weyl point will measure the same monopole charge as when $f\equiv 1$. An analogous
statement can be made about the $\pi$ Berry phase of the Dirac points of graphene in the 
presence of Coulomb repulsion. 
  
{\bf Orthogonal Weyl semimetals}: 
We present a case where the above characterization of   
a Weyl-like liquid using $\mc H_t$ breaks down. The idea being that particular interactions 
can induce a phase where the charge carrying quasiparticles have the properties of a WSM
but are \emph{orthogonal} to the electron due to fractionalization. 
Such a phase admits a simple and stable slave-particle description: the electron operator $c_{r,\s}$ can be written 
as the product $f_{r,\s}\tau_r^x$ of a slave fermion $f_{r,\s}$ carrying the \emph{charge} (and spin), and a slave 
Ising pseudospin $\tau_r^x$. A $\mathbb Z_2$ gauge redundancy emerges because of the decomposition.  
In terms of these slave operators, a WSM results when the $f$-fermions form a WSM while 
the pseudospins are ordered.  
However, if they become disordered, an orthogonal WSM results for the electrons: The    
$f$-fermions constitute a Weyl liquid since the pseudospins and $\mathbb Z_2$ gauge field are gapped, 
but they are orthogonal to the 
electrons (the electronic quasiparticle weight vanishes). The resulting orthogonal WSM   
is a cousin phase of the orthogonal metal\cite{orthog}.   
It has qualitatively the same thermodynamic and transport\cite{hosur-rev,lundgren} properties as a Weyl liquid: $T^3$ heat capacity, quantum oscillations\cite{drew} and AQH response. However, the electron Green's function $G$ shows a hard ``Mott'' gap, thus no Weyl points.   
In this sense, the AQH response can no longer be obtained using $\mc H_t=-G(0,\b k)\inv$. Instead, 
one has to use the $f$-fermion Green's function. We thus have an instance where the  
adiabaticity relation to bare electrons breaks down, but where  
the topological Hamiltonian approach can be adapted by   
identifying the low-energy excitations. A similar situation will arise for other   
orthogonal states, such as orthogonal topological insulators\cite{orthog_ti}.   

{\bf Conclusion:} We have shown how to characterize interacting WSMs via the many-body Berry curvature 
(derived from the zero-frequency Green's function) 
allowing the identification of the monopole structure of the Weyl points.   
We have argued that the existence of quasiparticles is not necessary in this, for example the latter are marginally destroyed
in a WSM with long-ranged Coulomb repulsion. As a natural extension, we note that $\mc H_t$ can also be 
used to efficiently identify Weyl nodes 
lying away from the Fermi surface, for example in a doped Weyl semimetal, which proves much simpler than
resolving the full spectral function.  
In closing, our work shows the importance of the Berry connection derived from the Green's function 
in the study of correlated fermions, especially their robust (quasi)topological 
features, in the gapless regime. 
We have illustrated that these ideas can be implemented numerically to study realistic models.   
 
\emph{Acknowledgments}: WWK is particularly indebted to D.M.~Haldane, Y.B.~Kim, S.S.~Lee and 
T.~Senthil for discussions.  
We acknowledge stimulating exchanges with A.~Go, B.I.~Halperin, J.~Maciejko, E.G.~Moon, 
R.~Nandkishore, 
A.~Vishwanath. WWK is grateful for the hospitality of
Harvard and the Princeton Center for Theoretical Physics, where some of the work was completed.
MK\ns was supported by the Austrian Science Fund (FWF) Project No.\ts J 3361-N20.   
Research at Perimeter Institute is supported by the Government of Canada through Industry Canada and by the Province of Ontario through the Ministry of Research and Innovation.   

\onecolumngrid
\appendix
\tableofcontents

\section{Anomalous quantum Hall conductivity via Green's functions and Berry curvature}
\label{ap:ahe}
\subsection{Green's function expression for anomalous Hall conductivity} 
\begin{figure}
        \centering
        \includegraphics[width=0.18\textwidth]{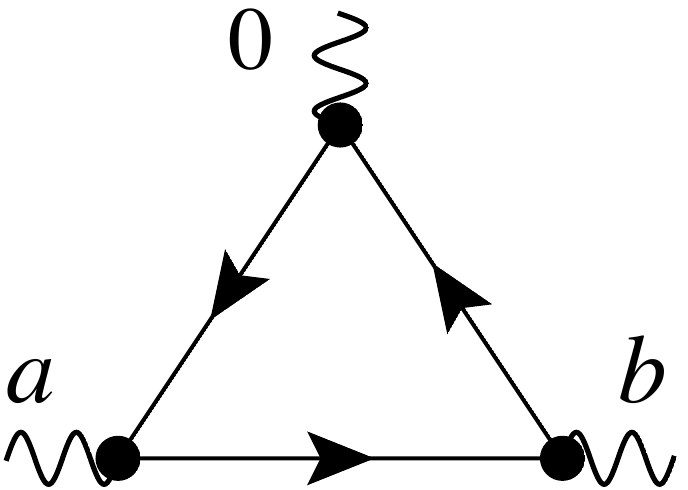} 
        \caption{ \label{fig:abj}
         Adler-Bell-Jackiw anomaly type Feynman diagram used to evaluate the DC anomalous quantum Hall response, \req{aqh}. 
         The external energy-momenta are set to zero; 
         $a,b$ are spatial indices (set to $y,z$, respectively, to get $\s_{yz}$). The fermion lines
         correspond to exact Green's functions $G(k)$, while the disks to the exact vertices $\G_\mu(k,k)=\pd_\mu G(k)\inv$.}
\end{figure}

To obtain the AQH conductivity $\s_{ab}$, we use the Kubo formula. The time-reversal odd part of the
current two-point (polarization) function that is relevant for $\s_{ab}$ reads:
\begin{align} \label{eq:Pi-gen} 
  \Pi_{ab}(q)= \frac{1}{(2\pi)^2} \e_{ab\rho\la}q^\rho K^\la + \cdots \,,
\end{align}
where $(\cdots)$ refers to terms unimportant for the DC AQH response. The Roman/Greek indices run over 
spatial/spacetime dimensions. In contrast to the rest of the paper, we use real time and
frequency in this subsection. 
Note that we have not
assumed Lorentz invariance in \req{Pi-gen}; rather the above form is dictated by current conservation, which
implies $q^\mu\Pi_{\mu\nu}(q)=q^\nu\Pi_{\mu\nu}(q)=0$. We have set $e=\hbar=1$; the  
later choice explains the extra factor of $1/2\pi$ compared to Eq.\ts(2) of the main text. $q=(q^0,\b q)$ is a four-momentum
corresponding to the external electromagnetic perturbation, and $K^\la$ is a $q$-independent four-vector. 
In the Kubo formula for the conductivity, 
$\s_{ab}=\lim_{q^0\ra 0}\lim_{\b q\ra\b 0}\Pi_{ab}(q)/q^0$, we first need to take the limit $\b q=\b 0$. 
We can thus set $\b q=\b 0$ in the
above, which fixes the index $\rho=0$. Further, we can assume without loss of generality that $\b K=(K_x,0,0)$ is along 
the $x$-direction. We thus obtain  
\begin{align}
  \s_{yz}=\frac{K_x}{(2\pi)^2} = \frac{\pd}{\pd q^0}\Pi_{yz}(q^0,\b 0)\big|_{q^0\ra 0} \,. \label{eq:Pi_der}
\end{align}
Now, the exact polarization function reads  
\begin{align}
  \Pi_{yz}(q)= \int \frac{dk^0 d\b k}{(2\pi)^4}\Tr\left[ \G_y^{(0)}(k,k+q) G(k+q)\G_z(k+q,k)G(k)\right] \,,
  \label{eq:pi}
\end{align}
where $\G_c/\G_c^{(0)}$ denotes the irreducible vertex function of the interacting/non-interacting theory. 
($\G_\mu(k,k')$ has incoming fermion energy-momentum $k$, and outgoing one $k'$.) 
Taking the frequency-derivative of \req{pi}, and using the Ward identity\cite{schrieffer,mahan} associated with 
charge conservation (which is valid on the lattice),  
\begin{align}
  \G_\mu(k,k) = \pd_\mu G(k)\inv\,, \label{eq:ward}
\end{align}
leads to the desired result for the anomalous quantum Hall response $\s_{yz}$: 
\begin{align}
  \s_{yz}=\frac{1}{96\pi^4} \int_{-\infty}^\infty d\w\int d\b k \; \e_{\mu\nu\rho x} 
   \Tr[G(\pd_\mu G\inv)G(\pd_\nu G\inv)G(\pd_\rho G\inv)] \,, \label{eq:aqh}
\end{align}
where we are using $e=\hbar=1$. 
The corresponding ``triangle'' Feynman diagram is shown in \rfig{abj}, where the external legs are
at zero energy and momentum. The frequency derivative of the polarization function inserts an external
photon line, leaving behind a three-point function. 
We note that the above formal manipulations are an extension to 3+1D of 
the corresponding ones in 2+1D used to obtain the quantized Hall conductivity of 
an interacting quantum Hall state\cite{so,ishikawa_86} using the exact Green's function.  
The result was anticipated in Ref.\ts\onlinecite{ahe_haldane} for Fermi liquids. In fact,
the above derivation is general, and not specific to interacting WSMs.

It is straightforward to obtain \req{aqh} for a system of free-fermions, starting with the Kubo formula, \req{Pi_der}.   
Indeed, minimal coupling the fermions 
to an external vector potential $\b A$, $\mc H(\b k)\ra \mc H(\b k + \b A)$, gives the
following expression for the spatial vertices: $\G_c(k,k)=-\pd_c\mc H(\b k)$. The time (or energy) component of
the vertex is simply the identity matrix, $\G_0(k,k)=1$, because the scalar potential $A_0$ couples to the fermion 
density. Note that the vertex function satisfies the Ward identity \req{ward}, where $G(k)\inv=k^0 -\mc H(\b k)$. 

\subsection{From Green's functions to Berry curvature}

We explicitly derive the expression for the Berry curvature in terms of the
Green's function of a system of non-interacting fermions:
\begin{align}
 b_c(\b k)= (\nabla\times \b a)_c = \frac{1}{3!}\int_{-\infty}^\infty \frac{d\w}{2\pi} \e_{\mu\nu\rho c}
  \Tr\left[G(\pd_\mu G\inv)G(\pd_\nu G\inv)G(\pd_\rho G\inv)\right] \,, \label{eq:B_using_G} 
\end{align}
where $\b a(\b k)$ is the Berry connection. In most of what follows, we suppress the $\b k$-dependence to lighten the notation.
Crucially, the above expression can be applied for the topological Hamiltonian $\mc H_t=-G(0,\b k)\inv$, and the associated 
topological Green's function, $G_t(i\w,\b k)\inv=i\w-\mc H_t(\b k)$, to recover the Berry 
curvature of the \emph{interacting} Weyl liquid. 
In that case the LHS of \req{B_using_G} is replaced by the generalized Berry flux $(\nabla\times \b{{\mc A}})_c$.  

We begin with a lattice system of free fermions defined by the Bloch Hamiltonian $\mc H(\b k)$, which includes
the chemical potential shift. Again, this covers the topological Hamiltonian $\mc H_t$. 
Its band structure is $\{\xi_n(\b k)\}$. Consider a $\b k$-point
away from the Fermi surface, such that the occupied levels, $\xi_n(\b k)<0$, are separated
from the unoccupied ones ($>0$) by a gap. Let us derive \req{B_using_G} for the $x$-component
of the Berry flux density, $b_x$. As can be easily checked, the $3!$ non-trivial combinations of the indices $\mu,\nu,\rho$
give the same answer. Let us thus pick $(\mu,\nu,\rho)=(y,z,0)$, such that 
\begin{align} 
  b_x = \int \frac{d\w}{2\pi}  
  \Tr\left[G(\pd_y G\inv)G(\pd_z G\inv)G(\pd_0 G\inv)\right] \,, \label{eq:Bx}   
\end{align}
since the fully anti-symmetric tensor evaluates to $\e_{yz0x}=1$.
Since the free Green's function reads $G(i\w,\b k)\inv=i\w-\mc H(\b k)$,
we obtain:
\begin{align}
  \pd_0 G\inv = i\,; \qquad
  \pd_c G\inv = -\pd_c\mc H\,. 
\end{align}
Note that these are the vertices of the non-interacting theory at vanishing momentum transfer, as
discussed in the previous subsection. 
Using these relations, together with an orthonormal set of Bloch states, $\mc H(\b k)\ket{n\b k}=\xi_n(\b k)\ket{n\b k}$,
we obtain 
\begin{align}
  b_x &= i\sum_{m,n}\bra{n\b k}\pd_y \mc H\ket{m\b k}\bra{m\b k}\pd_z\mc H\ket{n\b k} 
  \int \frac{d\w}{2\pi}\frac{1}{(i\w-\xi_n)^2(i\w-\xi_m)}\,; \\
   &=i\sum_{m,n} \frac{\Theta(-\xi_m\xi_n)\sgn(-\xi_m)}{(\xi_m-\xi_n)^2} 
  \bra{n\b k}\pd_y \mc H\ket{m\b k}\bra{m\b k}\pd_z\mc H\ket{n\b k} \,,
\end{align}
where the second equality follows after performing the $\w$-integral by contour integration. 
We have introduced the step function: $\Theta(x)=1$ when $x>0$ and vanishes for $x<0$. 
It constrains $\xi_m,\xi_n$ to have opposite signs. Note that $\xi_n(\b k)$ does not vanish since
$\b k$ was chosen away from the Fermi surface. Now, by taking derivatives of the matrix elements
$\langle n \b k|m\b k\rangle$ and $\bra{n \b k}\mc H\ket{m\b k}$, we get 2 relations that will
allow us to simplify the above expression:
\begin{align}
  0 &= (\pd_a\bra{n\b k})\ket{m\b k} +\bra{n\b k}\pd_a\ket{m\b k}\,, && \kern-4em \forall m,n \\ 
  \bra{n \b k}\pd_a\mc H\ket{m\b k} &=-\xi_m(\pd_a\bra{n\b k})\ket{m\b k}-\xi_n\bra{n\b k}\pd_a\ket{m\b k}\,, && \kern-4em m\neq n 
\end{align}
Using these we arrive at
\begin{align}
  b_x=-i\sum_{m,n} \Theta(-\xi_m\xi_n)\sgn(-\xi_m)\bra{n\b k}\pd_y\ket{m\b k}\bra{m\b k}\pd_z\ket{n\b k}
\end{align}
Moving the derivatives around and making use of the completeness relation $\sum_n\ket{n\b k}\bra{n\b k}=1$
to eliminate one of the summation variables, we
get the desired result:
\begin{align}
  b_x=-i\sum_{\xi_n<0}\left[ (\pd_y\bra{n\b k})\pd_z\ket{n\b k}- (y\leftrightarrow z)\right]\,,
\end{align}
which can be readily checked  to be equal to $(\nabla\times\b a)_x$. 

We note that the above derivation connecting the Berry curvature $b_x$ to the 
frequency integral of the ``triangle trace'', \req{Bx}, also holds in two dimensions.
Both in two and three spatial dimensions, the suitably normalized integral of $b_x(\b k)$ over the spatial momentum of 
yields $\s_{yz}$. 

\section{Cluster Perturbation Theory}
\label{ap:cpt}
Cluster perturbation theory\cite{senechal} (CPT) can be understood as embedding  
an exactly solvable reference system in the physical system. In particular, 
we consider a cluster decomposition of the physical lattice  
as our reference system. 
The cluster Green's function $\t G$ can be evaluated 
using exact diagonalization and naturally depends on the 
$N_c$ cluster sites. This Green's function can be written in Lehmann representation
\begin{align}
 \t G(i \omega) = Q \frac{1}{i \omega - \Lambda} Q^\dag \;.
 \label{eq:lehmann}
\end{align}
In this notation particle and hole excitations are combined, $\Lambda$ is a diagonal matrix
which gives the locations of the poles, and the matrix $Q$ determines their weights. 

The Green's function of the physical system $G$ is then obtained from strong-coupling 
perturbation theory 
\begin{align}
 G=\t G + G T \t G \,,
\end{align}
where $T$ describes the intercluster hopping. Using the Lehmann representation of the
reference Green's function \req{lehmann} and the Fourier transform 
$v_{\b k}^\dag=\frac{1}{\sqrt{N_c}}(e^{-i\b k \cdot \b r_1},e^{-i\b k\cdot \b r_2},\ldots,e^{-i\b k\cdot \b r_{N_c}})$, we   
obtain for the Green's function of the physical system
\begin{align}
 G(i \omega,\b k) = v_{\b k}^\dag Q \frac{1}{i \omega-(\Lambda +Q^\dag T Q)}Q^\dag v_{\b k} \;.
\end{align}
Therefore, the topological Hamiltonian is given by
\begin{align}
 \mc H_t(\b k)= \left[v_{\b k}^\dag Q \frac{1}{\Lambda +Q^\dag T Q}Q^\dag v_{\b k}\right]^{-1}\;.
\end{align}
In the models we consider, $\mc H_t(\b k)$ is a matrix in spin and sublattice space (the latter, for model II only). 
Diagonalizing it gives the topological band structure $\t\xi_n(\b k)$. 

CPT becomes exact in the limit $U \to 0$ as well as  $U \to \infty$, and is controlled in
the sense that convergence can be monitored by increasing the cluster size and
with that the quality of the self-energy of the reference system. 
\begin{figure}
        \centering
        \includegraphics[width=0.5\textwidth]{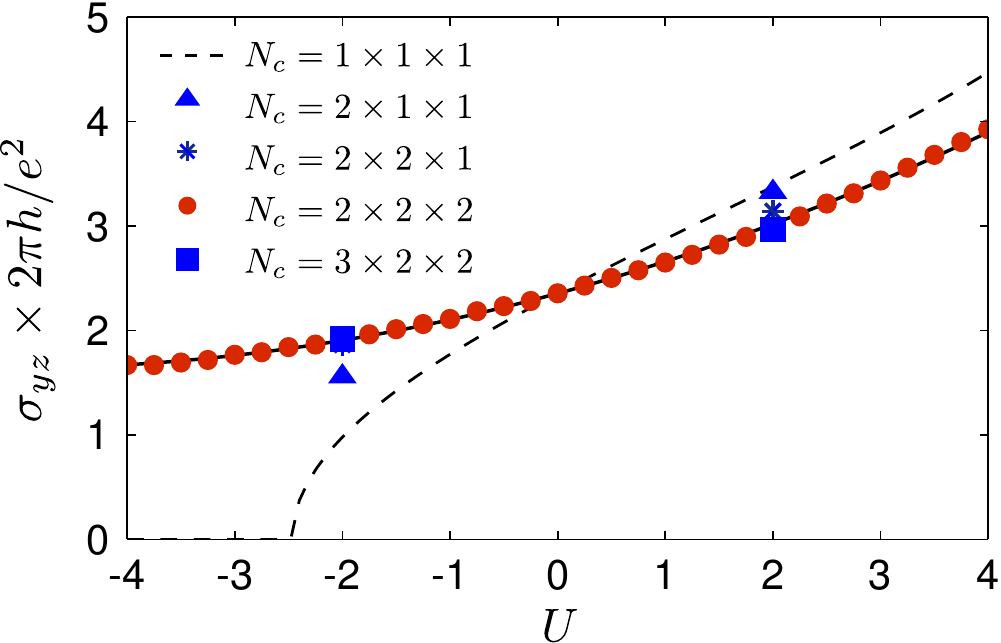} 
        \caption{ \label{fig:conv}
         A counterpart of Fig.\ts1b of the main text showing the dependence of the anomalous Hall response
       of a WSM (model I) on the Hubbard $U$. We have shown the results for different cluster sizes, $N_c$, to illustrate that reasonable
     convergence has been achieved already for $N_c=2\times2\times2$.}  
\end{figure} 

As an example, in \rfig{conv}, we illustrate the convergence of the anomalous Hall response of model I with respect to
the cluster size $N_c$ (same parameters as for Fig.\ts1b of the main text). It can be seen that the results for
$N_c=2\times2\times2$ and $N_c=3\times2\times2$ can hardly be distinguished. Further numerical refinement due to an increased
$N_c$ would be very costly in terms of computational resources and would not alter our results. This conclusion is reasonable
given that the CPT results are well bounded by those from the mean field and strong coupling calculations, as can be seen
in Fig.\ts1b of the main text.

\section{Weyl nodes of Models I and II}
\label{ap:models}
\subsection{Model I}
The tight binding part of Model I was introduced in Ref.\ts\onlinecite{ran}. Below we review the non-interacting
WSM band structure, and analyze the effects of the Hubbard term on the Weyl fermions using strong and weak coupling
expansions.

\textbf{Non-interacting band structure: }Diagonalizing the non-interacting Bloch Hamiltonian, 
Eq.\ts(5) of the main text, defines the band structure
\begin{equation}
 \xi_n(\b k) = \pm \left[{\lbrace 2 (\cos k_x-\cos k_0) +m (2-\cos k_y -\cos k_z) \rbrace^2 + 4 \sin^2 k_y
+ 4 \sin^2 k_z }\right]^{1/2}\;.
\end{equation}
The hopping amplitude $t$ has been set to unity.
We have also set the chemical potential to zero, as appropriate for half-filling. 
The non-interacting Weyl nodes are then obtained from the zeros of $\xi_n(\b k)$:
\begin{subequations}
\begin{alignat*}{3}
 \b k &=(\pm k_0,0,0)&&\\
 \b k &= (\pm\cos\inv(\cos k_0-m),0,\pi) \qquad &&\text{if} \quad |\cos k_0-m| \leq 1\\
 \b k &= \left(\pm\cos\inv(\cos k_0-m),\pi,0\right) \qquad &&\text{if} \quad |\cos k_0-m| \leq 1\\
 \b k &= (\pm\cos\inv(\cos k_0-2m),\pi,\pi) \qquad &&\text{if} \quad |\cos k_0-2m| \leq 1\;.
\end{alignat*}
\end{subequations} 
When $m>1$, the Hamiltonian has the minimal allowed number of Weyl nodes, \ie only two.
This is the case we focus on. 
The Hamiltonian of model I breaks TRS, since $\Theta H \Theta^{-1} \neq H$, where $\Theta = (-i \sigma^y) \hat K$ ($\hat K$ is the complex conjugation operator).  

As discussed in the main text, short ranged attractive or repulsive interactions in a WSM are irrelevant in 
the renormalization group sense. Therefore,
the Weyl nodes, which are hedgehogs of the Berry curvature, cannot be destroyed by local interactions 
but their position in the BZ can, and generically will, change.  
In the following, we discuss perturbative estimates for the renormalization of the Weyl fermions induced 
by local interactions in both the strong and weak coupling limits. 

\textbf{Strong-coupling limit: }Starting with the Hamiltonian of model I in the atomic limit ($t=0$) 
\begin{equation}
  H_{\rm a} = U n_\uparrow n_\downarrow -\mu (n_\uparrow + n_\downarrow) +  \underbrace{2(m-\cos k_0)}_{\equiv M} c^\dag \sigma_x c  \; .
\end{equation}
we perturbatively turn on the hopping to neighboring sites. Using the equations of motion technique, or alternatively
the Lehmann representation, we first calculate 
the atomic limit Green's functions at half-filling, $\mu=U/2$,  
\begin{align}
 G_{\rm a,\uparrow \uparrow}(\omega) = G_{\rm a,\downarrow \downarrow}= \frac{\omega}{\omega^2 -  (M+\frac{U}{2})^2} \qquad G_{\rm a,\downarrow \uparrow}(\omega) =G_{\rm a,\uparrow \downarrow}= \frac{M + \frac{U}{2}}{\omega^2 -  (M+\frac{U}{2})^2}\;.
\end{align}
Second, we set up the topological Hamiltonian, $\mc H_t(\b k) = -G^{-1}_{\rm a}(0)+T(\b k)$, 
using first order perturbation theory in $t/U$. 
Here, $T(\b k)$ is the hopping matrix, \ie the $\b k$-dependent part of the Bloch Hamiltonian $\mc H(\b k)$.
(The $\b k$-independent part has already been included in $H_{\rm a}$.)   
Diagonalizing $\mc H_t(\b k)$ gives the effective band structure (setting $t=1$)
\begin{equation}
 \tilde \xi_n(\b k) = \pm \left[{ \left\lbrace2 (\cos k_x-\cos k_0)+ m(2-\cos k_y - \cos k_z) +{U}/{2}\right\rbrace^2+4 \sin^2 k_y + 4 \sin^2 k_z}\right]^{1/2} ,
\end{equation}
from which we obtain the interacting Weyl points
\begin{subequations}
\begin{alignat*}{3}
 \b k &=(\pm \cos\inv(\cos k_0-{U}/{4 }),0,0) \qquad &&\text{if} \quad |\cos k_0-{U}/{4 }| \leq 1\\
 \b k &= (\pm\cos\inv(\cos k_0-m-{U}/{4}),0,\pi) \qquad &&\text{if} \quad |\cos k_0-m-{U}/{4}| \leq 1\\
 \b k &= (\pm\cos\inv(\cos k_0-m-{U}/{4}),\pi,0) \qquad &&\text{if} \quad |\cos k_0-m-{U}/{4}| \leq 1\\
 \b k &= (\pm\cos\inv(\cos k_0-2m-{U}/{4}),\pi,\pi) \qquad &&\text{if} \quad |\cos k_0-2m-{U}/{4}| \leq 1\;.
\end{alignat*}
\end{subequations}  
The functional dependence of the Weyl points on $U$ in the strong-coupling limit is shown in Fig.\ts1b of
the main text, in the parameter regime where only two Weyl nodes are present. Recall that in that case
$\s_{yz}\propto 2k_0$, where $\pm\b k_0$ are the locations of the interacting Weyl nodes.    

\textbf{Weak-coupling limit: }The renormalization of the Weyl nodes in the weak-coupling limit $U\ll t$,
can be obtained from a variational Hartree-Fock calculation. We consider the case of half-filling for
which the interaction decouples as 
\[
 U n_{r,\uparrow} n_{r,\downarrow} \to \frac{U}{2} (n_{r,\uparrow} + n_{r,\downarrow}) - U m_x  c_r^\dag \sigma_x c_r +U m_x^2 \;,
\]
where we introduced the magnetization  $m_x=\frac{1}{2}\langle c_r^\dag \sigma_x c_r \rangle$ as an order parameter.
In the weak-coupling limit we thus obtain 
\begin{equation}
 H_\text{HF}= \sum_{\b k} m_x^2 U +c_{\b k}^\dag \big[\{2t(\cos k_x-\cos k_0)-U m_x +m(2-\cos k_y-\cos k_z)\}\s_x \\ +2t\sin k_y\,\s_y+2t\sin k_z\,\s_z\big]c_{\b k} \;.
\end{equation}
Minimizing the ground state energy, determines the optimal variational order parameter $m_x$ 
from which we can determine the position of the renormalized Weyl nodes, see Fig.\ts1b of the main text. 
 
\subsection{Model II}  
We introduce a new model for a WSM that respects time-reversal symmetry, which 
thus belongs to the other family of WSMs (as opposed to model I). It is defined on the cubic lattice. 

\textbf{Non-interacting band structure: }Model II has a charge-density-wave order determined by
$H_{\rm cdw} = \e \sum_{\b k} c^\dag_{\b k} c_{\b k+\b Q}$, with 
ordering wavevector $\b Q=(\pi,\pi,0)$. Therefore, the chemical potential in the $xy$ plane is 
staggered by $\pm \e$. In order to  
solve for the non-interacting band structure of Hamiltonian, Eq.\ts(6) of the main text, we double the unit cell and rotate it by $\pi/4$ in the $xy$ plane such that it includes 2 sites with local potentials $\e$ and $-\e$, respectively. The tight binding Hamiltonian thus reads 
\begin{equation}
 H_0 =\sum_{\b k} \begin{pmatrix}	 
              c_{\b k}^\dag & d_{\b k}^\dag
             \end{pmatrix}
             \begin{pmatrix}	
              \e + 2 \sigma_z \sin k_z & 2 \sigma_x \sin k_+ +2 \sigma_y \sin k_-  \\ 
              2 \sigma_x \sin k_+ +2 \sigma_y \sin k_- &-\e + 2 \sigma_z \sin k_z
             \end{pmatrix}
             \begin{pmatrix}	
              c_{\b k} \\ d_{\b k}
             \end{pmatrix}\;,
\end{equation} 
where $c_{\b k}/d_{\b k}$ is associated with the $\pm\e$ sublattice. We have defined the inplane momenta $k_\pm=k_X\pm k_Y$,
where $k_{X,Y}$ correspond to the enlarged unit cell obtained when $\e\neq 0$. 
The associated non-interacting band structure consists of the 4 bands:
\begin{align*}
 \xi_n(\b k) &= \pm \left[4 \sin^2 k_++ 4 \sin^2 k_- + (\e+2 \sin k_z)^2\right]^{1/2} \\
 \xi_n(\b k) &= \pm \left[4 \sin^2 k_++ 4 \sin^2 k_- + (\e-2 \sin k_z)^2\right]^{1/2} \;.
\end{align*}
This band structure defines a WSM with 16 Weyl nodes located at a combination of any
\begin{align*}
 k_X & \in \{0,\pi\}\;, \qquad k_Y  \in \{0,\pi\}\;, \qquad k_z  \in \{\pm \sin\inv \e/2,\pi \pm \sin\inv \e/2 \}\;.
\end{align*}

The Hamiltonian of model II preserves TRS, $\Theta H \Theta^{-1}=H$. However, 
inversion symmetry $P H(\b k) P^{-1}=H(- \b k)$ is explicitly broken.   

\section{Lifetime of Weyl and Dirac excitations}
\label{ap:liquid}
We provide a brief analysis of the lifetime of the nodal excitations in
Weyl and Dirac liquids in $d$ spatial dimensions. We also discuss the marginal liquid case that arises with long-range Coulomb repulsion.
First, the Dirac/Weyl liquid states are obtained by considering 
linearly dispersing nodal fermions interacting with short-ranged interactions. For $d\geq 2$,
such interactions, both repulsive and attractive, are irrelevant in the renormalization group sense.
Indeed, a coupling $U$ parameterizing a four-fermion contact term (say of density-density type $\rho(x)^2$) scales like
$[U]=\w^{d-1}$, where $\w$ is the real frequency (or energy). Thus when $d>1$,  
$U$ vanishes as $\w^{d-1}$ at low energy $\w\ra 0$.
Therefore, the $U$-driven scattering rate of the Weyl or Dirac excitations in the liquid can be obtained
from a perturbative scaling analysis of
the self-energy: 
\begin{align}
 \g(\w) &\equiv \im\S_R(\w+i0^+,\b k_0)\,, \\
 &=\w (U_{\rm eff}/\La)^2 \; \sim \; |\w|^{2d-1}\,,
\end{align} 
where $\b k_0$ corresponds to the nodal point, and $\La$ is a UV energy scale such as the bandwidth. 
The overall factor of $\w$ in the second equality arises on dimensional grounds, 
while $U_{\rm eff}$ is the effective running coupling 
constant describing the short-range interaction. It appears squared due to the perturbative interaction
where a fermion creates a single virtual particle-hole pair.
From the discussion above, we have $U_{\rm eff}(\w)\sim |\w|^{d-1}$,
yielding the energy-dependent scattering rate $\w^{2d-1}$. This confirms that for $d>1$, the excitations
become sharp as $\w\ra 0$. In other words, an infinitely lived quasiparticle emerges at the node. 
In the case of Weyl or Dirac liquids in $d=3$, we obtain $\g(\w)\sim |\w|^5$. 
For $d=2$, we recover the standard result for two-dimensional Dirac liquids, such as graphene
with short-range interactions: $\g\sim \w^3$. 

In the presence of the $1/r$ Coulomb repulsion, the Weyl/Dirac liquid breaks down marginally.
Indeed, in both two and three dimensions the coupling parameterizing the Coulomb $1/r$ interaction in the action,
$V\int dt d^d\b x d^{d}\b y \rho(x)\rho(y)/|\b x-\b y|$, is \emph{marginal}. In other words,
$V_{\rm eff}\sim \w^0$, so that the scattering rate becomes $\g \sim \w V_{\rm eff}^2\sim \w$, implying a marginal destruction of the nodal quasiparticle
in both $d=2$\cite{sarma} and $3$\cite{abrikosov}.   

%

\end{document}